\documentclass[twocolumn]{article}

\usepackage{tikz}
\usetikzlibrary{arrows.meta,positioning,calc,decorations.pathreplacing,backgrounds}
\usepackage{subfigure}
\usepackage[T2A]{fontenc}
\usepackage{CJKutf8}
\usepackage{url}
\usepackage[colorlinks=true, allcolors=blue]{hyperref}
\usepackage[
    backend=biber, 
    natbib=true,
    style=apa,
    sorting=nyt
]{biblatex}
\usepackage{float}
\usepackage[section]{placeins}
\usepackage{graphicx}%
\usepackage{multirow}%
\usepackage{amsmath,amssymb,amsfonts}%
\usepackage{amsthm}%
\usepackage{mathrsfs}%
\usepackage[title]{appendix}%
\usepackage{xcolor}%
\usepackage{textcomp}%
\usepackage{manyfoot}%
\usepackage{booktabs}%
\usepackage{algorithm}%
\usepackage{algorithmicx}%
\usepackage{algpseudocode}%
\usepackage{listings}%

\theoremstyle{thmstyleone}%
\theoremstyle{thmstyletwo}%

\theoremstyle{thmstylethree}%
\title{Just Another Hour on TikTok: ID sampling to obtain a complete slice of TikTok}
\author{
  Benjamin Steel\\
  McGill University\\
  \texttt{benjamin.steel@mail.mcgill.ca}
  \and
  Miriam Schirmer\\
  Northwestern University\\
  School of Communication\\
  \texttt{miriam.schirmer@northwestern.edu}
  \and
  Derek Ruths\\
  McGill University\\
  \texttt{derek.ruths@mcgill.ca}
  \and
  Juergen Pfeffer\\
  Technical University of Munich\\
  \texttt{juergen.pfeffer@tum.de}
}
\date{}

\begin{document}









\maketitle

\abstract{TikTok is now a massive platform, and has a deep impact on global events. Despite preliminary studies, issues remain in determining fundamental characteristics of the platform. We develop a method to extract a representative sample of >99\% of posts from a given time range on TikTok, and use it to collect all posts from a full hour on the platform, alongside all posts from a single minute from each hour of a day. Through this, we obtain post metadata, video media, and comments from a close-to-complete slice of TikTok, and report the critical statistics of the platform. Notably, we estimate a total of 269 million posts produced on the day we looked at, that 18\% of videos on the platform feature children, and that at least 0.5\% of posts contain artificial intelligence-generated content.}



\section{Introduction}

TikTok has reshaped global culture through its algorithm-driven platform that promotes short-form video content. Due to its immense reach \parencite{pew2024social}, especially among younger voters \parencite{huang2022genz,leppert2024americans}, and its ability to amplify content quickly through its algorithm, politicians, advocacy groups, and individuals use TikTok to share political messages, mobilize voters, and raise awareness about issues \parencite{mackinnon2021examining,medina2020dancing}. 

However, concerns have been raised about the spread of misinformation \parencite{allen2022misinformation}, algorithmic bias \parencite{karizat2021algorithmic}, lack of transparency in how political content is managed \parencite{ghaffary2023behind}, and child safety on the platform \parencite{soriano2023tiktok}. Additionally, TikTok's ownership and data practices have sparked debates over national security and foreign influence, particularly regarding its potential use in shaping public opinion during elections \parencite{european2024commission, spring2024tiktok, finkelstein2025information}. At the same time, little is known about the platform’s inner workings and its global activity level. To gain insight here, we need good data from the platform.

Data collection methods on TikTok are limited \parencite{steel2024invasion}, and do not allow for representative sampling. Search mechanisms on the site are limited to hashtag based search which stops at 1000 returns, fuzzy search, and user- and song-based search \parencite{steel2024invasion}. All of these methods limit the extent to which a true complete slice of the platform can be examined. While TikTok offers a Researcher API to selected scholars \parencite{corso2024we}, it has been established recently that this API shows significant deviations in terms of metadata and availability of posts when compared to the TikTok website \parencite{pearson2024beyond}. Additional limitations of the API include restrictive user agreements \parencite{bakcoleman2023tiktok} and the inability to return posts originating from Canada \footnote{https://developers.tiktok.com/doc/research-api-codebook/}. We must obtain stronger sampling methods if we are to gain contextualized insight into a platform of TikTok's magnitude \parencite{pfeffer2023just}. Without representative samples, researchers cannot properly contextualize or validate findings about behaviour on TikTok.

To address this, we developed a novel method for collecting >99\% of posts uploaded within a timespan, enabling the collection of a close-to-complete slice of TikTok for the first time. We did this by determining how TikTok creates IDs for its posts, in order to create an ID generator that can allow us to search for all possible posts, which allows us to bypass the insufficient available search elements, as done previously on YouTube \parencite{zhou2011counting}. We expect this method and extensions of it to be invaluable to research efforts to understand the scope, dynamics, and impact of TikTok.

We used this method to collect near-complete samples of TikTok. Considering prevailing research themes around TikTok and other social media platforms, we provide insight on the following points:

\begin{itemize}
    \item \textbf{The total volume of content on TikTok}: Estimating this allows us to understand the true size of the platform for the first time, and therefore understand its total usage, a critical task to avoid sampling bias \parencite{ruths2014social}.
    \item \textbf{Distributions of post content on TikTok}: We present data on country post origin, engagement statistics, and artificial-intelligence generated content from our representative dataset. This allows us to understand global usage and consumption patterns, critical when considering the disparity between global populations and the TikTok user base \parencite{mislove2011understanding}.
    \item \textbf{Media content}: Video media content is the central medium of TikTok, and understanding video topic breakdowns is therefore critical for insight into what role topics have on the platform. Additionally, TikTok has proven incredibly popular for children \parencite{ofcom2024children}, raising safety concerns, so understanding the role of children in content on the platform is essential.
    \item \textbf{Comment statistics}: Comment volume, comment posting time, and comment language. TikTok comments give us insight into mass user engagement patterns.
\end{itemize}

In addition to releasing the code needed to reproduce this work, we also release fundamental statistics regarding these phenomena. We expect these will be useful to a wide-array of future studies: both for the studies themselves and to calibrate their findings against the broader backdrop of TikTok behavior. We invite researchers to approach us for specific data requests.

\section{Results}

We developed a method to collect >99\% of TikTok's global post content, and used it to collect posts made between 5-6pm UTC on the 10th of April, 2024 (an arbitrary, mundane date, to obtain baseline data), and posts from the $42^{nd}$ minute of each hour from 10pm UTC on the $9^{th}$ of April, till 10pm on the $10^{th}$ of April, to obtain an across-day representative sample. We first present the high-level statistics of our two datasets in Table \ref{tab:tiktok-stats}.

\begin{table}[htbp]
\small
\centering
\begin{tabular}{p{34mm}|p{13mm}|p{15mm}}
\toprule
\textbf{\newline Metric} & \textbf{\newline 1 Hour} & \textbf{1 Minute\newline Each Hour} \\
\midrule
No. of Videos & 5,157,488 & 1,879,335 \\
No. of Unique Users & 3,715,672 & 1,760,335 \\
Mean Comment Count & 4.92 & 4.18 \\
Mean Like Count & 177.9 & 145.8 \\
Mean Share Count & 14.1 & 9.14 \\
Mean View Count & 2,533.2 & 2243.9\\
Mean Video Duration (s) & 20.69 & 20.42 \\
\bottomrule
\end{tabular}
\caption{Statistics from the two datasets we collected. The two datasets overlap at the $42^{nd}$ minute of the hour we looked at in this work, so total counts are lower than the sum of the two counts.}
\label{tab:tiktok-stats}
\end{table}

\subsection{Overall post volume}
\label{sec:post_volume}

\begin{figure}[htbp]
    \centering
    \subfigure[Posts per second.]{
        \includegraphics[width=0.4\textwidth]{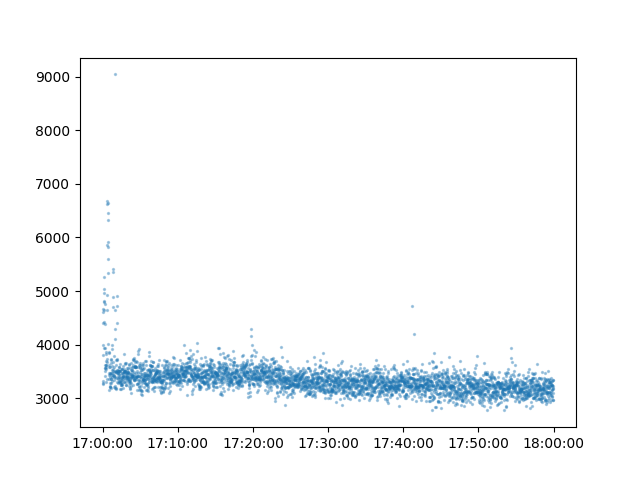}
        \label{fig:persec}
    }
    \subfigure[Posts per minute.]{
        \includegraphics[width=0.4\textwidth]{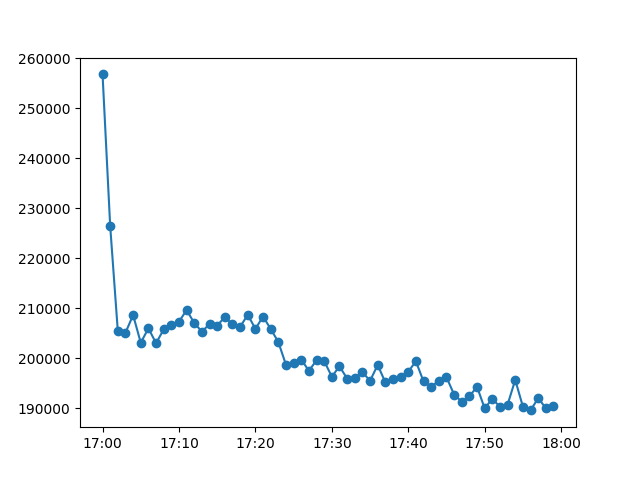}
        \label{fig:permin}
    }
    \subfigure[Estimated posts per hour.]{
        \includegraphics[width=0.4\textwidth]{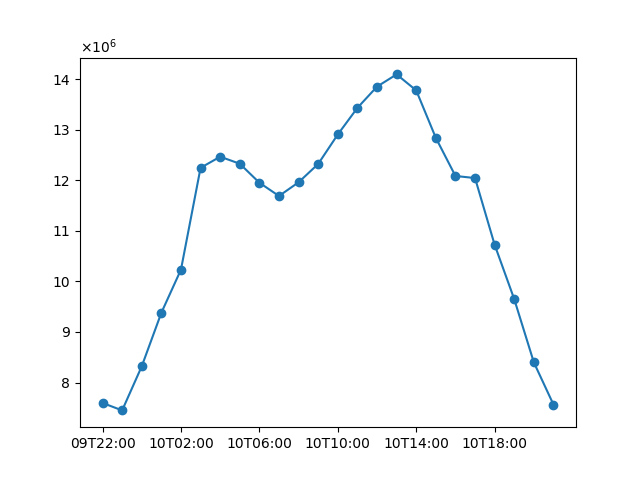}
        \label{fig:perhour}
    }
    \caption{TikTok posts per second and per minute on the 10th of April, 2024, 5pm--6pm UTC, as well as the estimated posts per hour over a 24 hour time period.}
    \label{fig:persecmin}
\end{figure}

The nature of our dataset offers us the most comprehensive assessment of the size of TikTok to date. For the observed hour, we collected 5 million posts. By looking at error messages for other requested posts, we determined that there were originally 12 million posts created in this hour (with many made inaccessible since post time, see Sec. \ref{sec:errors}), with a mean of 3.3k posts per second (min.-max. $2.8k-9.0k$, see Fig. \ref{fig:persec}). The aggregated numbers per minute range from $189,620$ to $256,743$ (mean $200,737.3$) posts per minute (Fig. \ref{fig:permin}). Here, because we want to consider all posts that were available in the original time-span, included no longer available posts, we use the create time that can be extracted from the post ID, see Section \ref{sec:id_generation}.

In order to estimate the posts per hour for an entire day, we collected the posts of the $42^{nd}$ minute of every hour. For our collected hour, the $42^{nd}$ minute is slightly less active ($\sim97.3\%$) than the mean of the other $59$ minutes. Consequently, we used this correction factor to calculate the estimated number of posts per hour. The result can be seen in Fig. \ref{fig:perhour} and shows an interesting two-peaked temporal pattern. These estimated hours aggregate to 269.3 million posts over a 24-hour time period.

We note that the time distribution of posts per hour looks markedly different from \citet{corso2024we}. We see that there are 28.5\% more posts in the first minute than in the mean of the other minutes, implying scheduled posts.

Of posts that are still available when we collected data, and therefore that we have metadata for, we see that there is a 14\% higher posting volume at the 0th second of each minute, implying scheduled posts, or bot activity. View counts for posts posted at the zeroth second get significantly more views (p-value: 7e-5), with an mean of 3,899 views at the zeroth second vs. 2,502 views for all other seconds. Videos at the 0th second more frequently use hashtags implying the desire to reach a broader audience, such as \textit{fyp}, implying that these posters are using post scheduling and hashtag use to attempt to reach a broader audience.

Understanding these characteristic statistics of the platform gives very valuable priors for future work on TikTok, and we hope they serve as such for the community.

\subsection{Engagement}

\begin{figure*}[htbp]
    \centering
    \subfigure[Distribution of comment count.]{
        \includegraphics[width=0.3\textwidth]{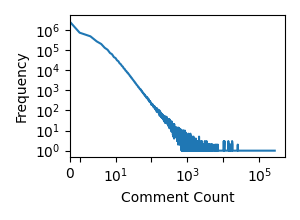}
        \label{fig:subfig1}
    }
    \subfigure[Distribution of view count.]{
        \includegraphics[width=0.3\textwidth]{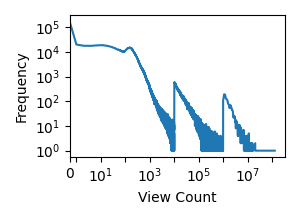}
        \label{fig:subfig2}
    }
    \subfigure[Distribution of like count.]{
        \includegraphics[width=0.3\textwidth]{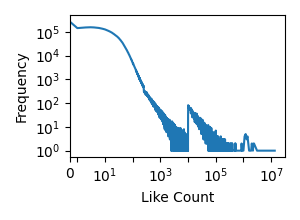}
        \label{fig:subfig3}
    }
    
    \subfigure[Distribution of share count.]{
        \includegraphics[width=0.3\textwidth]{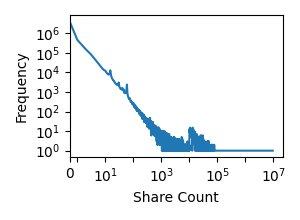}
        \label{fig:subfig4}
    }
    \subfigure[Distribution of video duration.]{
        \includegraphics[width=0.3\textwidth]{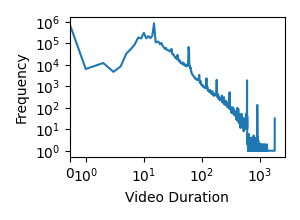}
        \label{fig:subfig5}
    }
    \caption{Histograms of various statistics in the post metadata, across the 1 hour dataset. We speculate that the peaks we see at round numbers on due to interval view count tracking optimization. However, they could also be evidence of inauthentic engagement.}
    \label{fig:video-stats}
\end{figure*}

Figure \ref{fig:video-stats} shows the distribution of post statistics from the 1 hour dataset we collected. While comment count and share count follow a typical exponential distribution as previously seen \parencite{corso2024we}, view count and like count show a mode at $\sim$100 views and $\sim$50 likes respectively, differing from the distributions reported by \citet{corso2024we}. This may be due to the large discrepancies between the Research API and the data returned to the platform interface demonstrated by \citet{pearson2024beyond}, highlighting the need to double-check Research API data with main platform data.

\subsection{Global usage}

\begin{figure*}[htbp]
    \centering
    \includegraphics[width=\textwidth]{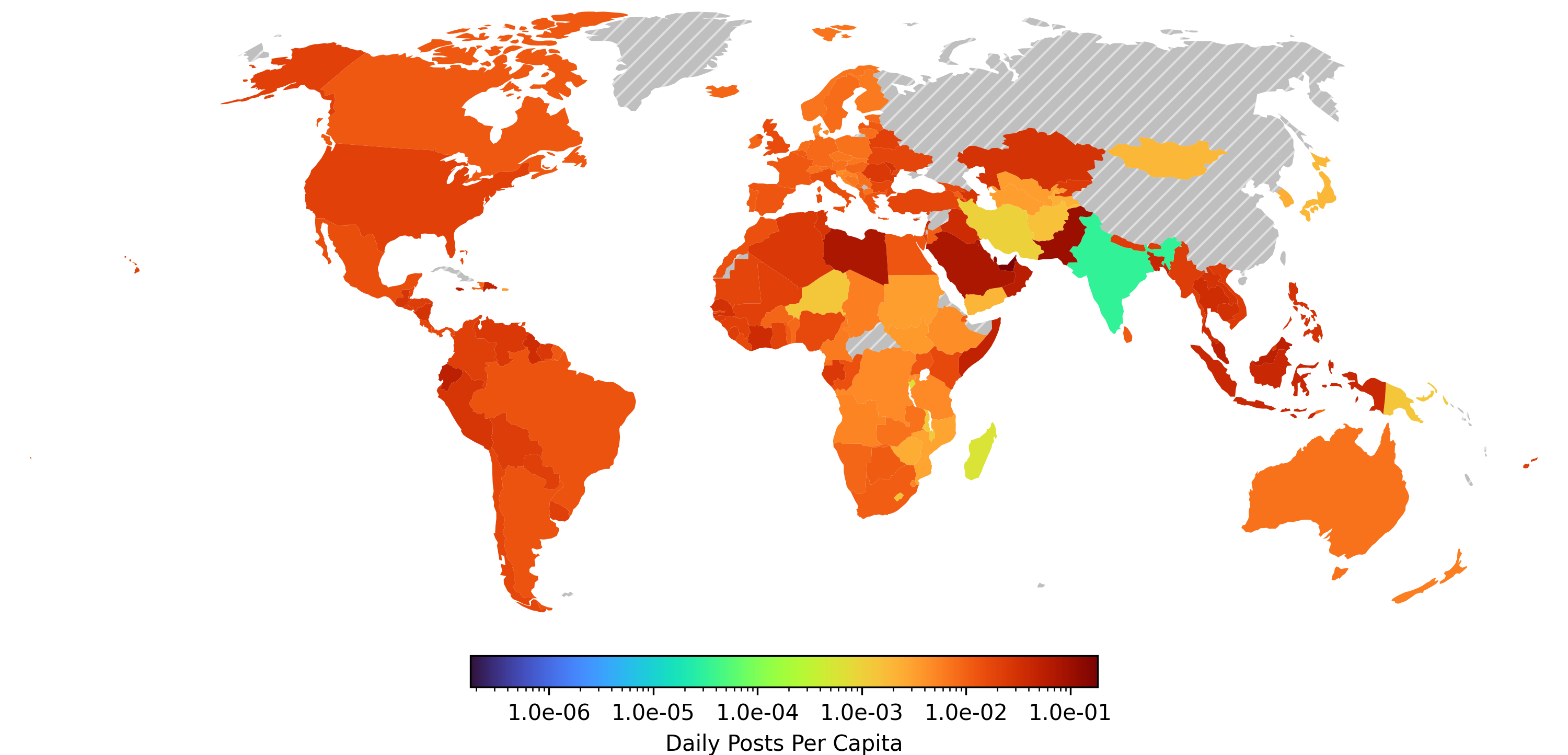}
    \caption{Choropleth of posting per capita for all countries, determined via the \textit{locationCreated} tag available on the post metadata. We normalize by the population of each country to show posts per capita, which specifically accounts for population rather than citizens \parencite{un2024world}. The counts are corrected for geographical regions that we estimate that we under-represent. Areas in grey are countries where we had less than 50 posts available.}
    \label{fig:country-count}
\end{figure*}

\begin{figure}[htbp]
    \centering
    \includegraphics[width=0.6\linewidth]{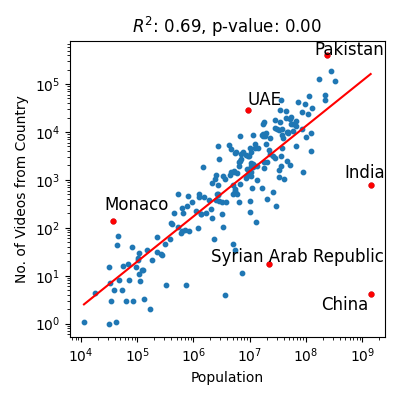}
    \caption{Regression of country population against post count in this dataset, with the 6 countries furthest from the trend line labelled, as determined by highest and lowest residuals from the linear regression. Red points correspond to the country name closest to them.}
    \label{fig:country-regression}
\end{figure}

We then looked at global platform usage, inferring post origin country from the \textit{locationCreated} attribute in the post metadata\footnote{https://developers.tiktok.com/doc/research-api-specs-query-videos}. While this indicator is likely confounded by VPN usage and other factors, it represents our best method for assessing a post's country of origin. As detailed later in Section \ref{sec:coverage}, our method has slight variation in regional representativeness, so we adjust counts based on known statistics from our method. We explain precisely how we do this in Section \ref{sec:usage_estimation}.

Figure \ref{fig:country-count} shows the popularity of TikTok across the world, correcting data by the population of the country \footnote{https://data.worldbank.org/indicator/SP.POP.TOTL}.

Figure \ref{fig:country-regression} shows a regression plot of country population against post count in our dataset ($R^2=0.69$, $p=0.00$). We see a strong fit to the data, evidence that we have successfully collected a globally representative dataset. We see that China, India, and Cuba, have the least posts in our dataset per capita, based on the residual of a fitted linear regression. We note that while TikTok was previously extremely popular in India \parencite{corso2024we}, there are now very few posts originating from India (as deemed by TikTok), likely due to the nationwide censorship policy on Chinese applications there \parencite{song2023can}. Of course, there could still be posts produced from India via virtual private networks (VPNs), which would be tagged with a different origin country by TikTok. TikTok is not officially available in China \parencite{nyt2024notiktok}, and Cuba has heavy internet restrictions \parencite{bbc2021cuba}. Monaco, Qatar, and the UAE, have the most posts in our dataset per capita.

\begin{figure*}[htbp]
    \centering
    \includegraphics[width=\linewidth]{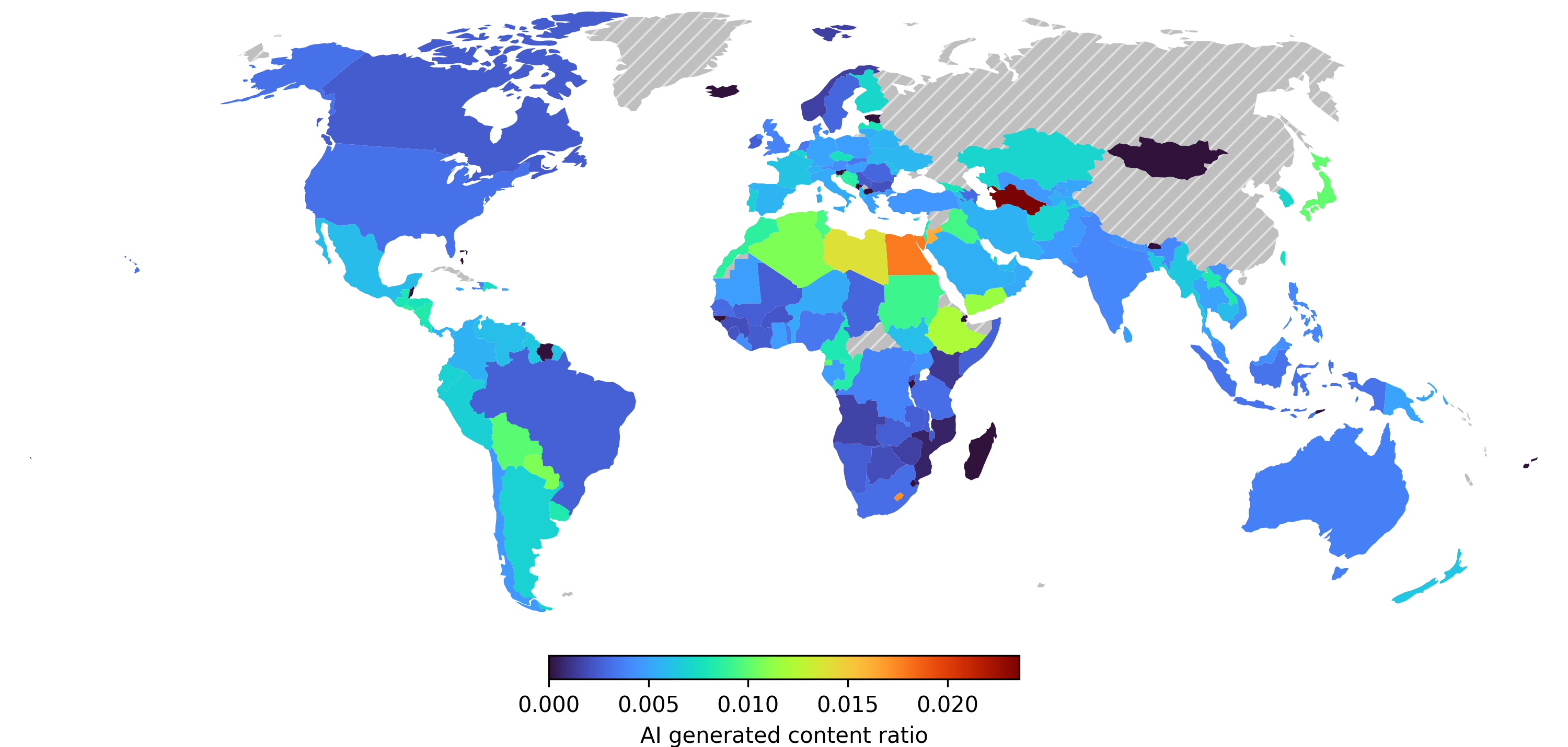}
    \caption{Share of posts tagged as AI generated content. Areas in grey are countries where we had less than 50 posts available.}
    \label{fig:aigc_map}
\end{figure*}

The post metadata also contains an attribute specifying whether the post media contains artificial-intelligence (AI)-generated content, whether tagged by the user or by the platform, including AI-generated images, video, audio, and filters \footnote{https://support.tiktok.com/en/using-tiktok/creating-videos/ai-generated-content}. In our 24-hour dataset, the percentage of content that is AI-generated according to TikTok is 0.5\%. This percentage varies by country (Fig. \ref{fig:aigc_map}). We find evidence of other posts containing AI-generated that are not tagged as such by TikTok, indicating the true statistic may be higher. However, we will leave a more thorough investigation of this important topic for future work.

\begin{figure}[htbp]
    \centering
    \includegraphics[width=0.6\linewidth]{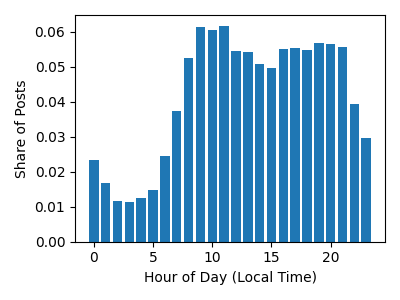}
    \caption{Distribution of local time posting frequency, found by looking at post time combined with the \textit{locationCreated} tag.}
    \label{fig:local_time}
\end{figure}

Using the \textit{locationCreated} attribute, we can also infer the distribution of when TikTok posts are created in the local time, see Figure \ref{fig:local_time}. In the case of countries spanning multiple time-zones, we use the mean hour from each time-zone associated with the country from the Olson time-zone database \footnote{https://github.com/stub42/pytz/blob/master/tz/zone.tab}. Inaccuracies are inevitable due to lack of sub-country location information, and countries spanning multiple time-zones, but the general trends should be preserved nonetheless.

TikTok's global popularity means that understanding TikTok's relative popularity across the world is necessary when trying to understand local aspects of the system, so we release country statistic breakdowns in our data release.

\subsection{Video Media}

TikTok is a video-based platform, and any understanding of the nature of the platform as a whole needs to come with a look at the actual video media data of the platform.

For each post in our dataset that has associated video media data (i.e. not a story-post), we downloaded the video media data. Seeing as we collected video media content after post collection on a rolling basis, we were not able to collect media content from posts that had been deleted in the intervening time. 

\begin{figure*}[htbp]
    \centering
    \includegraphics[width=\textwidth]{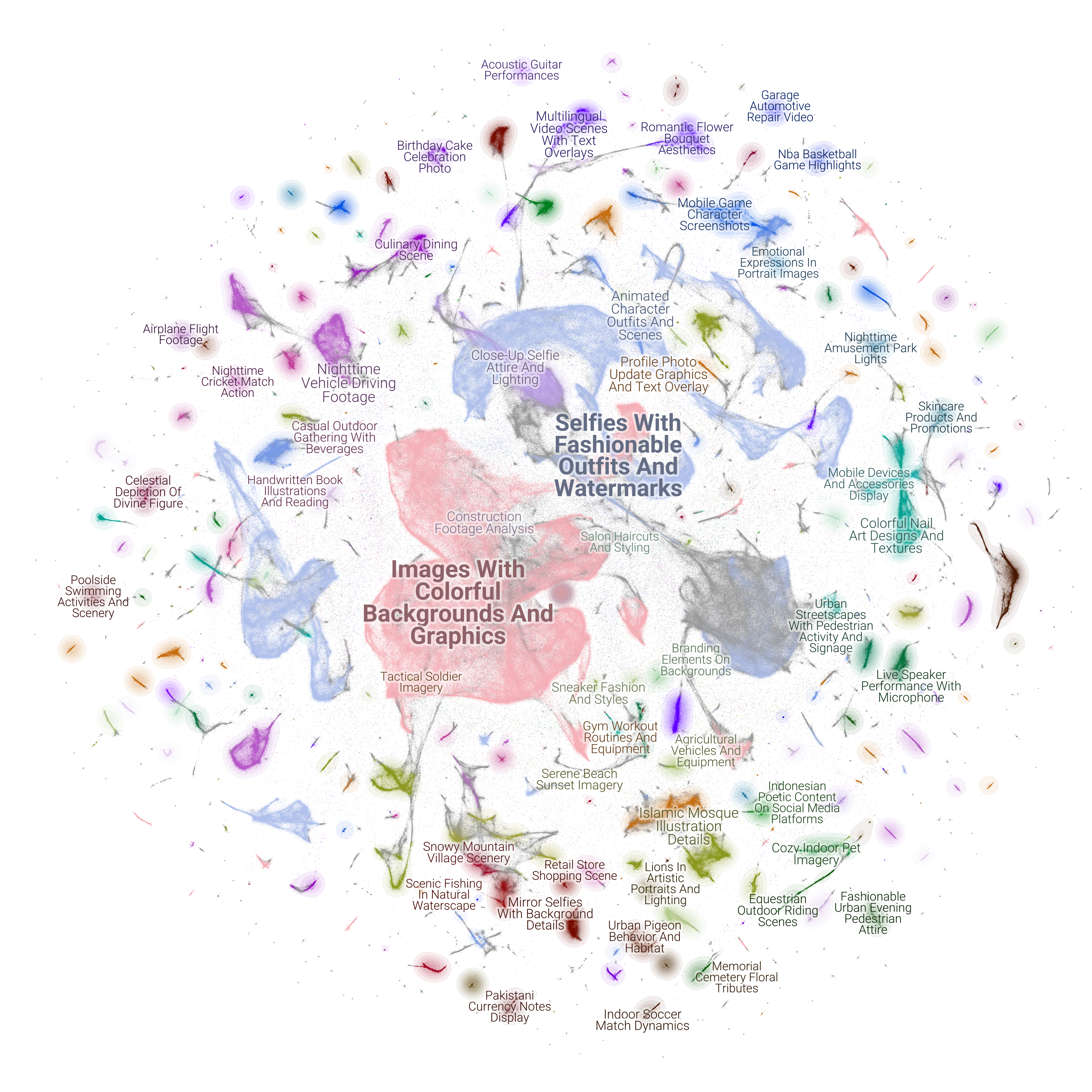}
    \caption{View of the topics covered by the posts, with each small dot representing a video, and distance between dots representing video content similarity. Common topics of videos are coloured and labelled with a description describing the topic.}
    \label{fig:topics}
\end{figure*}

We wanted to assess the kind of topics discussed on the platform. We, therefore, used a combination of topic modelling and vision language models to cluster the videos and describe each cluster. We plot the posts and clusters in Figure \ref{fig:topics}.

This analysis provides an exploratory glimpse at the range of topics touched on by TikTok content. We can see the broad array of these topics, with the largest being videos of a person talking directly to the camera, along with common activities in everyday life such as sports, transport, and lifestyle videos. In smaller clusters we find more uncommon topics, such as `Political Commentary on Police Operations in Pakistan' (0.025\% of videos), `Camouflaged Military Soldier Image' (0.06\%), and `Palestinian Flag Protest Imagery' (0.008\%).

The topic analysis completed here is not meant to be comprehensive. But by releasing the unsupervised topics discovered, we provide researchers an entry-point to contemplating the study and engagement of particular domains within TikTok.

\subsubsection{Presence of Children}

The use of TikTok by children has caused concern for safety on the platform \parencite{soriano2023tiktok}, but without knowing what percentage of content on the platform features children, we cannot know the true magnitude of the issue, or make connections between localised policy and content outcomes. 

We trained a classifier to detect if a video contains a child, with a mean F1 of 0.86 from 5-fold cross-validation on our train/test-set (precision 0.88, recall 0.81). We provide data, model selection, and training details in Section \ref{sec:child_classification}. We used this classifier to find that 18.00\% of the videos in our 24 hour dataset contain children (95\% confidence interval: [13.08\%, 24.62\%] from bootstrapped confidence intervals, see Section \ref{sec:child_classification} for details). This is a strikingly high statistic, and underscores how deeply embedded children are in platform activity. To the best of our knowledge, no comparable large-scale studies exist for TikTok or other platforms that would allow direct benchmarking or comparison of this figure.

\begin{figure*}[htbp]
    \centering
    \includegraphics[width=\textwidth]{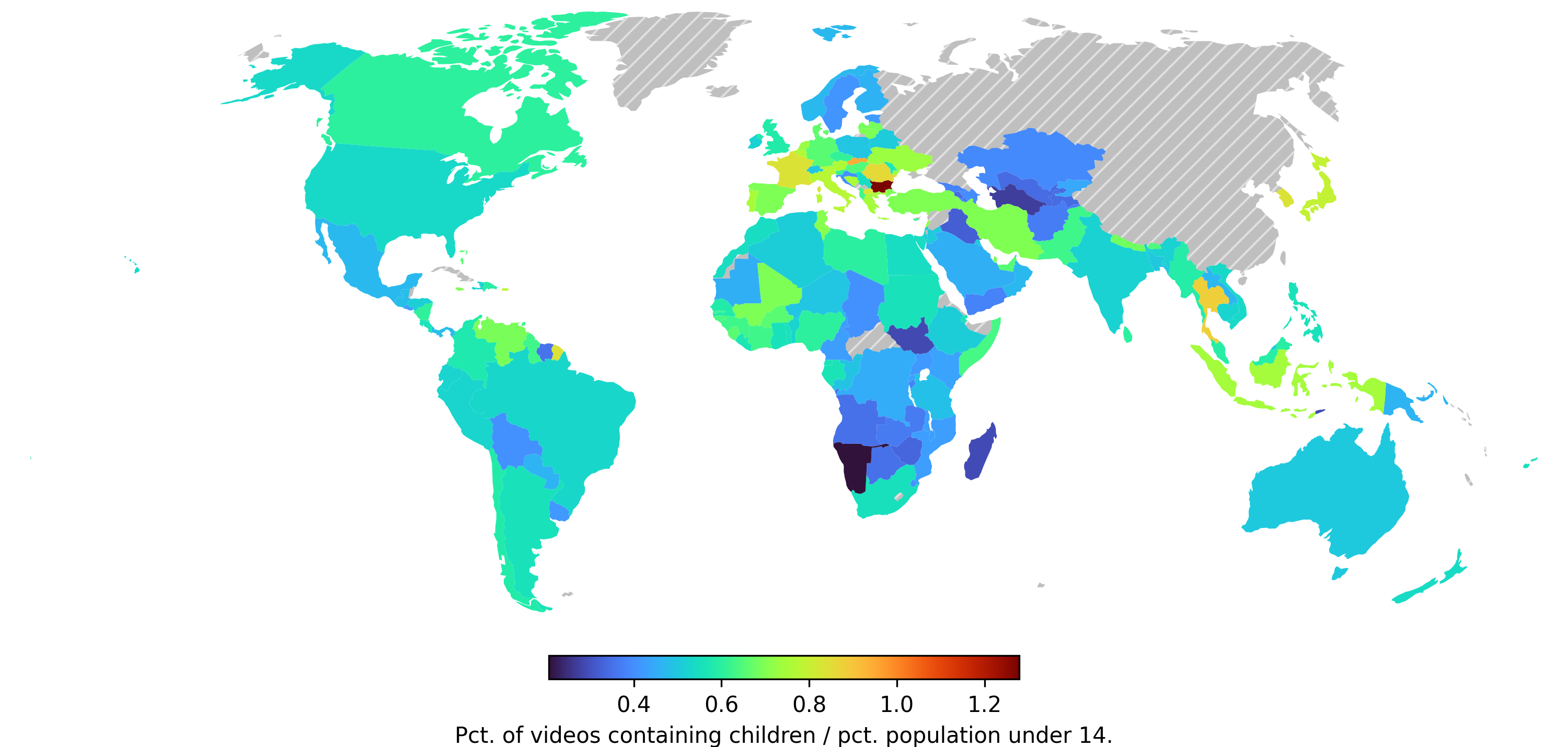}
    \caption{Choropleth of share of videos from our 1 minute per hour dataset containing one or more children, divided by the share of the population under the age of 14 \parencite{un2024world} (to avoid showing simply a map of countries with a youthful population), plotted by country. Areas in grey are countries where we had less than 50 posts or videos available.}
    \label{fig:child_map}
\end{figure*}

\begin{figure}[htbp]
    \centering
    \includegraphics[width=0.6\linewidth]{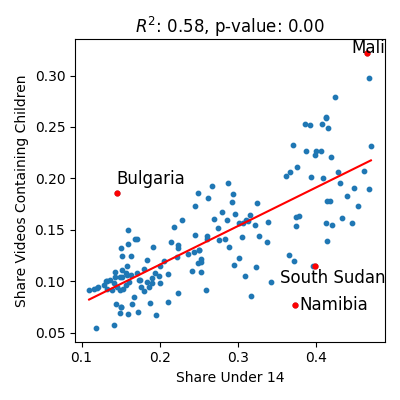}
    \caption{Scatter plot of countries, for each country showing the share of the population under the age of 14 against the share of videos from that country containing one or more children, as determined by our video classifier. We have removed countries where there was insufficient data simply by removing countries with below 50 posts in the dataset. We label the top 2 and bottom 2 countries with the highest residual from the linear regression, effectively showing countries with the most overrepresented and under-represented children in TikTok posts, respectively. Red points correspond to the country name closest to them.}
    \label{fig:child_regression}
\end{figure}

We observe variability in the fraction of videos containing children by country.  Figure \ref{fig:child_map} shows a choropleth of share of videos containing one or more children per country, adjusted by the share of the population under the age of 14 \footnote{https://population.un.org/wpp/} (to avoid showing simply a map of countries with a youthful population). We removed countries where there were less than 50 posts in the dataset. This effectively shows countries where there is a culture of posting videos containing children. Figure \ref{fig:child_regression} shows a scatter plot of countries comparing the share of the population under 14, and the share of videos containing one or more children - a linear trend is evident. We label the top 2 and bottom 2 countries with the highest residual from the linear regression ($R^2=0.51$, $p=0.00$), effectively showing countries with the most overrepresented and under-represented children in TikTok posts, respectively. The variation we observe may reflect broader differences in cultural norms around public sharing of children’s images, the popularity of family-centred content, differences in platform access and usage patterns, and differences in the implementation of child protection policies across countries.

With this analysis, we produced, to the best of our knowledge, the first attempt to measure what percentage of TikTok consists of content featuring children, lending context to our understanding of the role children play on TikTok.

\subsection{Comments}

Comments are an abundant form of mass engagement data on TikTok, providing language, semantic, and temporal data. Analysis of comments provides a rich lens through which we can understand TikTok's user base. 

We collected 18.4 million comments on the posts (excluding comments on comments), associated with 13.5 million commenters. We collected these comments after the collection process was complete, in September 2024, 5 months after the day the posts were posted, using Ensemble Data \footnote{https://ensembledata.com/}. The delayed comment collection process meant that we were not able to collect comments from posts that had been deleted since collection.

The median number of comments on the posts is 0, and the mean number is 4.92, with 51.9\% of posts receiving zero comments.

When looking at the distribution of comment creation time depending on second of the minute and minute of the hour, we see a flat distribution (std dev. across each second of a minute: 652, std. dev. across each minute of an hour: 3.7k, with a mean of 306k), indicating minimal scheduled posting of comments, and if automated comment posting is occurring, timing of posting is more sophisticated.

\begin{figure}[htbp]
    \centering
    \includegraphics[width=0.6\linewidth]{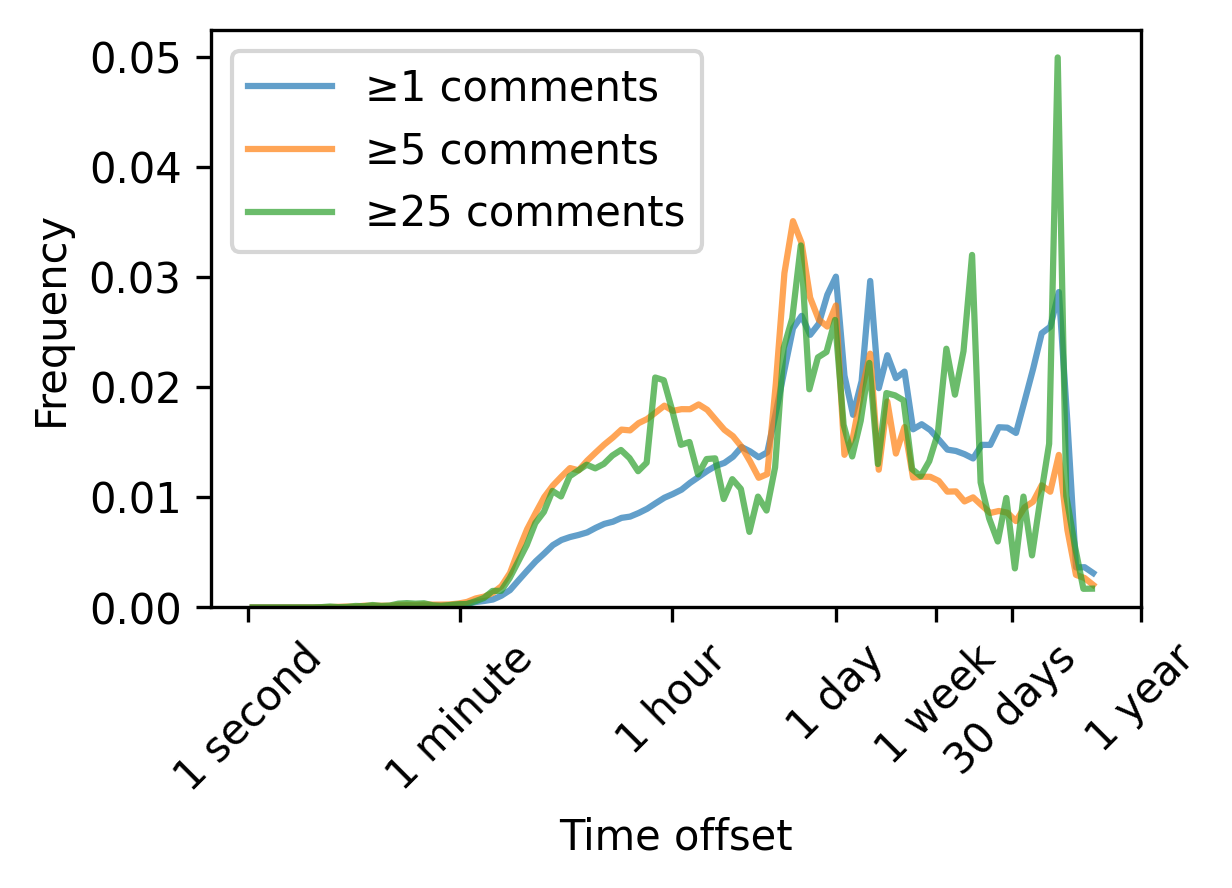}
    \caption{Time offset between post and comment creation by commenter frequency. We show offset distributions for authors who have at last 1 comment, authors who have at least 5 comments, and authors who have at least 25 comments, to show how commenting frequency affects commenting speed.}
    \label{fig:comment_offset}
\end{figure}

We also looked at mean offset between post and comment time, as shown in Figure \ref{fig:comment_offset}. We see that for all commenters, the modal time to comment is just under 1 day, and that comments from more frequent commenters ($\ge 5$ comments in our dataset) have a modal response delay of 1 hour from post create time. We can also see that around 2/3 months after a post is created, the commenting frequency on the post rapidly declines. This data shows us that while distribution processes on TikTok are fast, users are not engaging with the content until at least a few minutes past post time.

Looking at comment language, 47\% of comments are unknown (46\% of comments consist only of emojis, explaining this high percentage), 15\% are in English, 9\% are in Spanish, and 5\% are in Arabic. The high percentage of emoji-only commenting on TikTok could imply large quantities of inauthentic activity, as emojis would be an easy way to boost a post's engagement while requiring limited effort on behalf of the bot.

We will release the distributions from these analyses in our data release to allow other members of the community to use them to add context to their own work.

\subsection{Error Analysis}
\label{sec:errors}

A key feature of our collection method means that we receive a great number of error messages when we attempt to collect posts that either don't exist, or are not available for other reasons. To our knowledge, this has not previously been reported on and we found it informative of internal TikTok content management practices.

We received several different types of error message for a requested video, and we report the most common error types, excluding the error message `item doesn't exist', here, and in Table \ref{tab:errors}. We can see that an extremely large number of video posts are deleted ($\sim$23\%), or privatized (\textit{status\_self\_see}) ($\sim$23\%). We also gain insight into what seems to be the TikTok moderation process, with $\sim$7\% of videos unavailable at the time of collection due to a review stage (\textit{status\_reviewing}), 1.18\% of videos hidden after audit (\textit{status\_audit\_not\_pass}), and 0.16\% currently undergoing content classification for moderation (\textit{content\_classification}). We also see two error messages potentially related to geo-location, with \textit{cross\_border\_violation} (0.01\%) likely related to location-based content restrictions, for example when TikTok made new posts unavailable in Russia at the beginning of the 2022 invasion of Ukraine \parencite{tracking2022tiktok}.

\begin{table}[htbp]
\centering
\begin{tabular}{lrr}
\toprule
Status & Count & Pct. (\%) \\
\midrule
status\_deleted & 3,763,773 & 23.21 \\
status\_self\_see & 3,732,155 & 23.01 \\
status\_reviewing & 1,201,251 & 7.41 \\
status\_audit\_not\_pass & 191,690 & 1.18 \\
content\_classification & 26,654 & 0.16 \\
cross\_border\_violation & 2,134 & 0.01 \\
copyright\_geo\_filter & 296 & 0.00 \\
\bottomrule
\end{tabular}
\caption{Distribution of most common post error messages, excluding `item doesn't exist'. Reported percentage is out of all posts that at some point existed in our combined datasets. So from this we can infer that around 23\% of posts on TikTok are deleted. Posts can have multiple error status messages, but there is no overlap between posts with \textit{status\_self\_see} and \textit{status\_deleted}, so their similar count numbers seems to be a coincidence.}
\label{tab:errors}
\end{table}

When we look at deletion rates over different parts of our collection period, we see that of the posts that at some point existed (and either still exist, or taken down as evidenced by the error code), at 26 days after post time, deleted posts accounted for 12\% of posts, and at 215 days after post time, they accounted for 20\% of posts. TikTok reports only removing 1\% of posts \footnote{https://www.tiktok.com/transparency/en-us/community-guidelines-enforcement-2024-4}, indicating that these are largely user removals.

This data gives us a first insight into inaccessible content on TikTok, and future work can leverage these statistics to bring to light the moderation processes used by the platform, the rate of deletion, and the nature of cross-border content restrictions.

\subsection{Inauthentic Activity}

With any social media dataset, understanding human behaviour through it also requires understanding how much of the data is produced through inauthentic activity \parencite{davis2016botornot}. 

We know that bots exist on the platform \parencite{kolomeets2021analysis}, but there are no publicly available bot detection tools for TikTok. Furthermore, it is unlikely that existing bot detection methods transfer well \parencite{steel2024invasion}, due to the difference in text distribution and platform properties. In this dataset, we do not have full chronological data timelines to allow methods that exploit timestamps to work fully \parencite{mazza2019rtbust}, or network data to allow graph methods to work \parencite{jiang2016catching}.

Given these constraints, we used simple heuristics for bot detection \parencite{ferrara2016rise}. We find that there are 46 accounts with more than 100 posts each posted in the hour we looked at, and the top 5 highest frequency posting accounts, with between 1500 and 6000 posts in the hour we looked at, all have the default username of \textit{user} followed by 9 random digits. Their posts have an mean of 1.9 views each, but account for 0.28\% of the total 1 hour dataset. Looking at comments, there are 831 accounts that have at least 10 comments and whose mean commenting frequency is higher than 1 every 3 seconds, with most of their comments (88\%) consisting only of emojis. It's clear from these initial statistics that there is automated activity on the platform, but these methods can only produce a rough lower bound on the extent of automated activity on the platform. We therefore hope future work can narrow our bounds on the amount of inauthentic activity on the platform and in this dataset. 

\section{Discussion}


Altogether, the data in this work is critical for calibrating our understanding of TikTok, and ID sampling is critical to properly understand social platforms. We see substantial differences between outputs from our work, and the distributions available through the API \parencite{corso2024we} in Figures \ref{fig:perhour}, \ref{fig:subfig2}, and \ref{fig:subfig3}, further validating the need for API audits \parencite{pearson2024beyond}. We briefly illustrate the utility of the data with five exploratory case studies.

The invasion of Ukraine was touted as the `First TikTok War' \parencite{chayka2022watching}, and the way the invasion of Ukraine was covered on TikTok was investigated \parencite{primig2023remixing} as eyes turned towards the conflict. With this dataset, we can learn that while the mean view count of posts from Ukraine (4.1k) was higher than the global mean (2.2k), it is still lower than the mean view count of posts from the US (6.3k), quantifying our understanding of its global attention on the day posts for this work were collected. We can also show that, in keeping with TikTok policy, no posts are tagged with a Russian origin (although of course workarounds such as VPNs for posting from Russia are possible). 

In other places it's been established that there is a tendency in the global south towards engagement with content and culture from the global north \parencite{ozkula2024global}. Our data renders this in striking relief when we consider that the global south \parencite{world2025global} accounts for 81\% of posts but only receive mean view counts of 1.5k (compared to a baseline of 2.2k).

Finally, for those looking at topical issues on TikTok, our data provides an entry-point by way of topical clusters. For work on Palestinian activism on TikTok \parencite{cervi2023playful}, we can aid in initial investigations by showing that 410 posts in our hour dataset were clustered as a Palestine protest associated topic, but those posts had a significantly lower (p < 0.02) mean view count of 440 views compared to the hour dataset view mean of 2.5k. For work looking at the role of war on TikTok\parencite{kleisner2022tactical}, we can show that 2,995 posts in our hour dataset were clustered as a military/soldier associated topic, and those posts received an mean of 5.2k views. And for work looking at the link between TikTok and public health \parencite{kong2021tiktok}, we can add context by showing that 300 posts associated with the topic `Medical Examination and Healthcare Setting' had a significantly higher (p < 0.02) mean view mean of 28.8k views than the mean.

In order to allow others to use the data we have collected here to calibrate their studies, we have released aggregate data distributions. We understand that this release won't cover every use case, so we invite interested parties to contact us for additional specific statistics.

This dataset is not perfect. Collecting the data for this dataset took 5 months, meaning that the first batch of data was collected 2 months after the day chosen, and the last batch of data was collected 7 months after the day chosen. This introduces temporal bias into our dataset, where older posts will accumulate more engagement, and are also more likely to be taken down, or deleted. However, looking at our comment engagement offset (Fig. 6), we can see that the median time of engagement for all comment authors is around 29.9 hours after the video was posted. It seems likely that the median time from video post to comment engagement is similar to the median time offset for other engagement. This means that the vast majority of engagement with posts happens within two months, meaning that at least for engagement, the temporal bias of a collection process should have minimal effect. We also know that from looking at deletion rates at different minutes in our collection, there was an 8\% percentage point increase in deletion rates from the first minute of our hour to the last minute.

We accepted the slow collection rate of our method due to its ability to give us a full representative slice of TikTok. However, producing fully geographically unbiased statistics ideally requires a full 24 hour's worth of data is collected \parencite{pfeffer2023just}. This is why we collected our second dataset of a minute from each hour. But being able to see the full spectrum of activity would necessarily require a full 24 hours worth of data.

We also cannot speak to the proportion of authentic to inauthentic content in this dataset. Detecting bots and influence operations is a pressing issue on TikTok which inhibits researchers' abilities to study human vs. bot behavior on the platform. While we used simple heuristics to examine the least sophisticated types of bots on the platform, in the age of LLMs and advanced influence operations, more comprehensive methods must be used to be able to detect this activity, which we will leave for future work.

Despite these and other limitations, we believe the statistics we have presented here to be of substantive value to the research community. With this work we have presented the first large vertical slice of posts from TikTok, and delivered the essential statistics to help understand the platform. Our data release will allow those doing studies of TikTok to contextualize their findings, and work to deepen our understanding of this global platform.

\section{Methods}
\subsection{ID Generation}
\label{sec:id_generation}

Every entity (video, user, and comment) on TikTok has a unique ID attached, which takes the form of a very large integer. In order to create an ID generator, we need to know how these integers are created for every new video.

We have a starting point, which is that we know that if we convert the integer to a 64 bit binary representation, we can take the first 32 bits, convert these 32 bits back to an integer representation, we get a unix timestamp, which tends to be within seconds of the recorded creation time of the video recorded in the video metadata \parencite{Benson2020Tinkering} (Although for reasons we detail later, we believe this is specifically the time at which a post is created in the system, not when it is released, an important difference for scheduled posts, see Section \ref{sec:time_discrepancies}). This indicates that within the total 64 bits of the ID, there is encoded information. Many large scale distributed systems that need to generate globally unique IDs, use systems similar to the Snowflake system \parencite{Snowflake, Discord, Instagram, Mastodon}. If the ID generation method is similar to snowflake, then we would expect that the rest of the bits are some combination of milliseconds data, and data relating to TikTok's infrastructure (worker IDs, sharding IDs, sequence IDs etc.). If we can understand what composes the rest of these bits, when generating IDs, we can reduce the search space from $4.3 \times 10^{9}$ combinations down to a manageable number.

\citet{domingues2021analyzing} found that the $3^{rd}$ lowest hexadecimal value determines the type of the ID (post, user, comment etc.), meaning the $52^{nd}$ to the $56^{th}$ bits \parencite{domingues2021analyzing}.

We scraped a TikTok video metadata dataset of 225 thousand posts, attempting to capture posts that were likely produced in a variety of places to get high geographical distribution. The first method we used for this was datasets associated with accounts known to be in a specific country, which were available for Canada \parencite{pehlivan2025can} and Germany \parencite{bosch2021broken}. For countries where we were unable to find country specific seedlists, we collected posts using the hashtag for the name of the country in the country's official language. As long as a reasonable percentage of posts with the hashtag do indeed originate in that country, by looking at ID distribution, we can estimate whether a good ID coverage for that geographical region.

We then take the numerical IDs for those posts, and convert them to 64 bits of binary. We confirm that the first 32 bits of the ID correspond to the creation time of the video. This further implies that the TikTok ID generation system uses a Snowflake like method. We know that it isn't exactly using the snowflake method, as it uses the first bit as part of the timestamp (Snowflake reserves this first bit as always zero) \parencite{Snowflake}.  

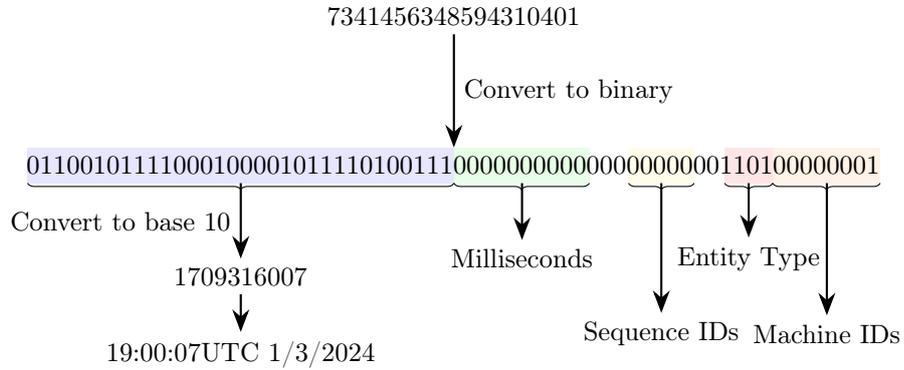
\begin{figure*}[htbp]
    \centering
    \begin{tikzpicture}[
        node distance=1.5cm,
        every node/.style={align=center},
        arrow/.style={-{Stealth[length=3mm]}, thick}
    ]
    \node (original) {7341456348594310401};

    \node[below=of original] (binary) {0110010111100010000101111010011100000000000000000000110100000001};
    \draw[arrow] (original) -- node[right] {Convert to binary} (binary);

    \begin{scope}[on background layer]
        \fill[blue!10] ($(binary.north west)!0.012!(binary.north east)$) rectangle ($(binary.south west)!0.5!(binary.south east)$);
        \fill[green!10] ($(binary.north west)!0.5!(binary.north east)$) rectangle ($(binary.south west)!0.655!(binary.south east)$);
        \fill[yellow!10] ($(binary.north west)!0.7!(binary.north east)$) rectangle ($(binary.south west)!0.775!(binary.south east)$);
        \fill[red!10] ($(binary.north west)!0.81!(binary.north east)$) rectangle ($(binary.south west)!0.865!(binary.south east)$);
        \fill[orange!10] ($(binary.north west)!0.865!(binary.north east)$) rectangle ($(binary.south west)!0.988!(binary.south east)$);
    \end{scope}

    \draw[decorate,decoration={brace, mirror}] 
        ($(binary.south west)!0.012!(binary.south east)$) -- 
        ($(binary.south west)!0.5!(binary.south east)$);

    \node[below=1cm of $(binary.south west)!0.256!(binary.south east)$] (base10) {1709316007};
    \draw[arrow] ($(binary.south west)!0.256!(binary.south east)$) -- node[left] {Convert to base 10} (base10);
    \node[below=0.5cm of base10] (timestamp) {19:00:07UTC 1/3/2024};
    \draw[arrow] (base10) -- (timestamp);

    \draw[decorate,decoration={brace, mirror}] 
        ($(binary.south west)!0.5!(binary.south east)$) -- 
        ($(binary.south west)!0.655!(binary.south east)$);
    \coordinate (ms_start) at ($(binary.south west)!0.578!(binary.south east)$);
    \node (ms_label) at ($(ms_start) + (0,-1cm)$) {Milliseconds};
    \draw[arrow] (ms_start) -- (ms_label);

    \draw[decorate,decoration={brace, mirror}] 
        ($(binary.south west)!0.7!(binary.south east)$) -- 
        ($(binary.south west)!0.775!(binary.south east)$);
    \coordinate (seq_start) at ($(binary.south west)!0.7375!(binary.south east)$);
    \node (seq_label) at ($(seq_start) + (0cm,-2cm)$) {Sequence IDs};
    \draw[arrow] (seq_start) -- (seq_label);

    \draw[decorate,decoration={brace, mirror}] 
        ($(binary.south west)!0.81!(binary.south east)$) -- 
        ($(binary.south west)!0.865!(binary.south east)$);
    \coordinate (prov_start) at ($(binary.south west)!0.8375!(binary.south east)$);
    \node (prov_label) at ($(prov_start) + (0cm,-1cm)$) {Entity Type};
    \draw[arrow] (prov_start) -- (prov_label);

    \draw[decorate,decoration={brace, mirror}] 
        ($(binary.south west)!0.865!(binary.south east)$) -- 
        ($(binary.south west)!0.988!(binary.south east)$);
    \coordinate (mach_start) at ($(binary.south west)!0.927!(binary.south east)$);
    \node (mach_label) at ($(mach_start) + (0cm,-2cm)$) {Machine IDs};
    \draw[arrow] (mach_start) -- (mach_label);

    \end{tikzpicture}
    \caption{Inferred ID composition}
    \label{fig:id-conversion}
\end{figure*}

To gather more information about the rest of the bits in the IDs, we found the inter-bit correlations by training logistic regression models to predict each of the bits in the IDs, given the rest of the bits as inputs. We found that bits 32 to 41 had no relation to each other, but that bits 42 to 63 had strong correlations, implying that there are a smaller set of unique values for the later part of this ID. Additionally, we see two clear sections in this later part, implying two different mechanisms for setting them, potentially worker IDs or sharding IDs.

\begin{figure}[htbp]
    \centering
    \subfigure[Value occurrence distribution for bits 32 to 42.]{
        \includegraphics[width=0.45\linewidth]{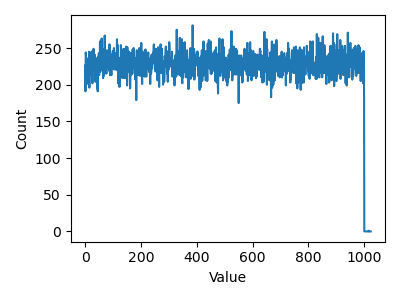}
        \label{fig:millisecond_bits}
    }
    \subfigure[Distribution of occurrence of values in bits 58 to 64. The different in distributions between geographic areas implies that this section of the ID encodes information related to the datacentre in which the ID was created, matching with the machine ID in Snowflake IDs \parencite{Snowflake}.]{
        \includegraphics[width=0.45\linewidth]{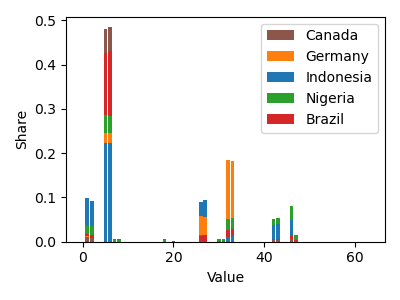}
        \label{fig:region_section}
    }
    
    \subfigure[Frequency of value occurrence in bits 45 to 50. There is minimal difference in the distributions between countries, implying this section of the ID is zone agnostic.]{
        \includegraphics[width=0.45\linewidth]{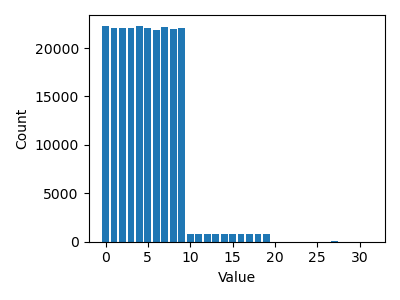}
        \label{fig:counter_section}
    }
    \subfigure[Visualization of the potential counter section of the ID, for a second of collected data, and two machine IDs.]{
        \includegraphics[width=0.45\linewidth]{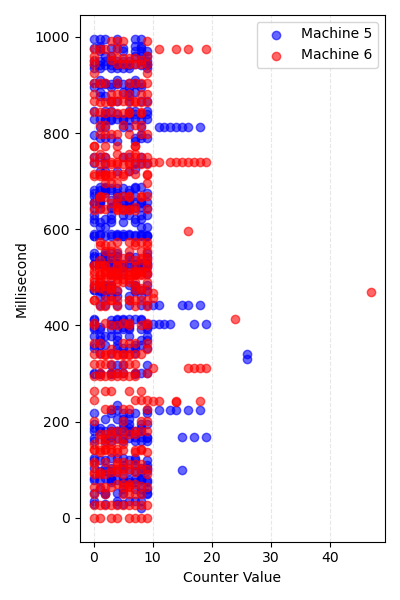}
        \label{fig:counters}
    }
    \caption{Representations of frequency of bits in the video IDs.}
    \label{fig:bits}
\end{figure}

We created a histogram of the first section, and found that values between 0 and 999 are roughly equally likely, with only 0.0004\% of values between 1000 to 1023, as shown in Figure \ref{fig:millisecond_bits}. This implies that this value is the millisecond section of the ID. The few other values are likely from other ID formatting mechanisms, or non video-type posts such as livestreams.


We also made histograms of the later section of the ID, and found that IDs were highly concentrated into specific values.

The original Snowflake ID system contains a machine specific ID in the generated ID, where there are a finite number of machines that generate new IDs. In a global distributed system, it's typically desired that different geographic regions (i.e. continents) are serviced by regional datacentres in order to reduce latency \parencite{carr2006google}, and therefore machine IDs will likely be region specific. For that reason, we collected datasets correlated with globally distributed countries. We used Canada and Germany to cover North America and Europe as we had pre-existing, large datasets of posts from those countries. We searched for large country specific pre-existing datasets for other countries, but to the best of our knowledge finding an absence of such datasets, collected datasets by getting posts using the hashtag of the name of those countries, with the intuition that most posts featuring those hashtags will have been created in those countries. Due to the hashtag search mechanism being limited to 1000 posts, we then collected posts from the accounts discovered via the hashtag method to obtain more posts.

To ensure we have a representative dataset, we wanted to create some statistical assurances that we have captured enough of the possible ID patterns that we can say with confidence we are capturing the vast majority of posts. To do this, we can turn to the Good-Turing estimator \parencite{gale1995good}, which, given our existing dataset of IDs, where we treat the non time encoding bits as the type of the ID, we can then estimate the number of each type of ID we have to estimate the probability of finding a previously undiscovered type of ID. This probability is then the probability of a video existing that we are not able to capture given our current technique. We report the corresponding percentage of coverage in Table \ref{tab:initial_dataset}.

\begin{table*}[htbp]
    \centering
    \begin{tabular}{c|c|c|c}
        \toprule
        Country & Method & No. posts & Estimated Coverage \\
        \midrule
        Canada & Seedlist & 118,994 & 99.93\% \\
        Germany & Seedlist & 35,256 & 99.94\% \\
        Nigeria & Country Name Search & 41,127 & 99.86\% \\
        Indonesia & Country Name Search & 19,509 & 99.83\% \\
        Brazil & Country Name Search & 14,079 & 99.77\% \\
        \bottomrule
        All & & 225,127 & 99.97\%
    \end{tabular}
    \caption{We estimate the percentage of posts we can capture, given the unique ID patterns we have found, using the Good-Turing estimator \parencite{gale1995good}.}
    \label{tab:initial_dataset}
\end{table*}

We see differences in the IDs in the later section as shown in Figure \ref{fig:region_section}. The difference in distributions between geographic areas implies that this section of the ID encodes information related to the datacentre in which the ID was created, which corresponds to the machine ID in Snowflake IDs \parencite{Snowflake}.


Finally we wanted to see if we could find sequence IDs, or counters, in the ID, as they are present in other distributed ID systems \parencite{Snowflake}. The only section that we have not yet established a reasonable inference of function for is bits 42 to 52. When comparing different ID types (video post, user, comment), we see differences in the distribution in these bits. However, there may also be a counter representation in this section. If found, and if they increment predictably, this would allow us to reduce the number of potential IDs that we need to try when doing ID generation. For video IDs, we see minimal variation in bits 42 to 52 other than in bits 45 to 50, for which we plot the distribution in Figure \ref{fig:counter_section}. 


We did find evidence of counters in the IDs, but they do not count up uniformly, sometimes missing steps, as shown in Figure \ref{fig:counters}. This means we cannot use them to narrow down our ID search space.


We then needed to generate these IDs, and request them from TikTok to see if they correspond to valid posts. To do this, we simply find all the possible variations of bit sequences for bits 42 to 63 (all the bits past the millisecond segment), and request each sequence of these for every millisecond. This results in 504 IDs to check for each millisecond.

\paragraph{Geographical Coverage}
\label{sec:coverage}

Due to there being datacentre specific sections of the ID, ensuring we have a good ID generation system is essential to having a globally representative collection of posts.

The Good-Turing estimator gives us confidence in what percentage of posts we have for each geographical area, but the question remains if we have good geographic coverage. After looking at the geographic distribution of the posts collected in the one minute from each hour dataset, we saw that the notable places where video counts were lower than others were East Asia, Oceania, India, and Russia. Though there are good reasons to believe that video counts in India and Russia would naturally be lower due to the ban of TikTok in those countries \parencite{song2023can,leave2023tiktok}, we collected extra datasets correlated with those regions (India, Australia, Japan, Russia) to check if there were previously undiscovered IDs, see details in Table \ref{tab:extra_countries}. We found that the unique IDs we had previously collected covered at least 99\% of the posts from those areas, and for India and Japan, at least 99.25\%, which we feel is sufficient to ensure a dataset is representative of those areas.

\begin{table*}[htbp]
    \centering
    \begin{tabular}{c|c|c|c|c}
        \toprule
        Country & Hashtag Name & No. Posts & Estimated Coverage & Fetch Overlap \\
        \midrule
        Russia & Россия & 10,288 & 99.35\% & 91\% \\
        Japan & \begin{CJK}{UTF8}{min}日本\end{CJK} & 9,363 & 99.30\% & 96\% \\
        Australia & australia & 6,522 & 99.13\% & 92\% \\
        India & india & 6,649 & 99.25\% & 97\% \\
        \bottomrule
    \end{tabular}
    \caption{We estimate the percentage of posts we can capture, given the unique ID patterns we have found, using the Good-Turing estimator \parencite{gale1995good}.}
    \label{tab:extra_countries}
\end{table*}

Specifically we notice that there are no posts attributed to Russia in our dataset, which is notable given that both India and Russia have banned TikTok, but we still obtain posts attributed to India. To ensure that we have not completed missed ID aspects that are associated with Russia, we collected the full metadata of the 9\% of posts we collected under the Russian hashtag that would not have been covered by our original fetch. We saw that none of them were attributed to Russia, only a range of other countries, implying that Russians using TikTok, or those ostensibly posting from Russia, are using VPNs, a common practice in other areas with banned internet services \parencite{chandel2019golden}.

\paragraph{Time Distribution Discrepancies}
\label{sec:time_discrepancies}

We noticed that when we look at post time distribution over each second of the minute using the \textit{createTime} attribute in the metadata, there is a 14\% increase at the $0^{th}$ second compared to the rest of the seconds, but that this distribution is flat when looking at the create time of the post as derived from the ID. As detailed in Section \ref{sec:post_volume}, these posts at the $0_{th}$ second are significantly different from posts in other seconds. We hypothesize that this means that the create time in the ID is the time when a scheduled post is created, and the \textit{createTime} attribute is the time when the video is made available as scheduled. This would imply that some of the posts we request are posts that are not yet available, and as such, would have an error message detailing that, which we hypothesize to be the \textit{status\_self\_see} error message.

\paragraph{Future Work}

We investigated the apparent counters found in the IDs, and trialled methods to predict the next valid video ID given previous counters. However, in the end we decided we wanted to get as complete a sample as we could, and therefore just requested all possible ID combinations. But if a method could be successfully developed here, it would greatly cut down the number of video IDs that would need to be requested, expanding the number of posts that could be collected through this method.

\begin{figure}[htbp]
    \centering
    \includegraphics[width=0.5\linewidth]{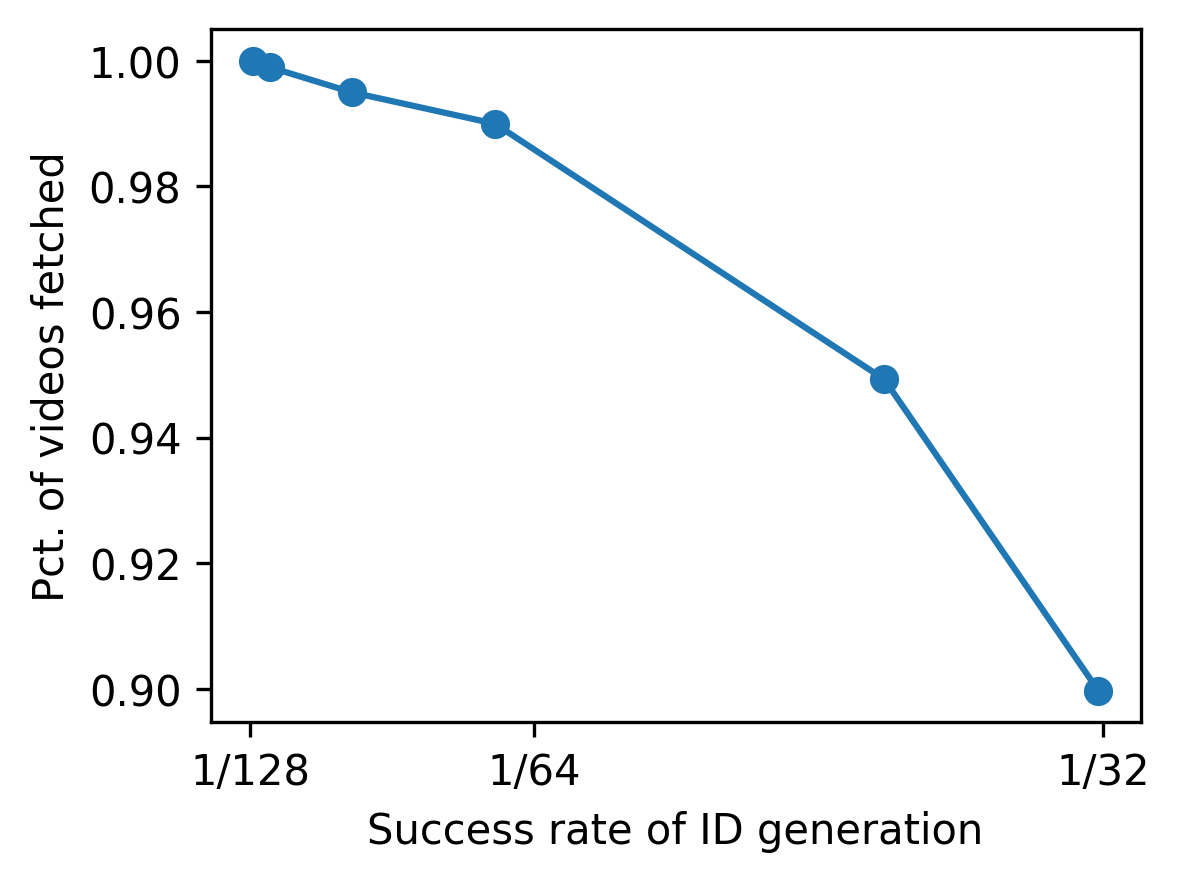}
    \caption{Success rate of ID generation vs percentage of posts successfully fetched out of all video we find for our combined dataset.}
    \label{fig:hit_rate_analysis}
\end{figure}

If we accepted a more probabilistic approach to video collection, we could have collected many more views for many fewer requests. However, this would have meant a less complete slice of TikTok. If from the data we have gathered, we can determine that ID patterns with more frequent video hits are not meaningfully different than ID patterns with less frequent occurrence of posts, we could confidently roll out such a system. Figure \ref{fig:hit_rate_analysis} shows the percentage of posts we can expect to get by aiming for higher hit rates.

We also looked at comment and user IDs, and a similar method used here could be applied to gather a sample of users. A random sample of comments would be more difficult due to the query for them requiring the video ID as well as the comment ID.

\subsection{Collection}

Once we could generate a full set of IDs, we needed to request them from TikTok. The typical way of viewing a TikTok video is by going to the URL: \textit{https://tiktok.com/@\{username\}/video/\{video\_id\}}. But we discovered that one can use an arbitrary username in the URL, and if the video corresponding to the video ID exists, one will be re-directed to the existing video page.

In total, 0.8\% of the IDs we requested returned either an available video, or an error indicating that the video had at some point existed.

We collected two datasets, one of all posts from 5-6pm UTC on the 10th April, 2024 (an arbitrary, mundane date, to obtain baseline data from the platform), and all posts from the $42^{nd}$ minute of each hour from 10pm UTC on the 9th April, till 10pm on the 10th April, to obtain across day representative data. The entire collection process took 5 months.

For the actual video collection process, we use Dask Distributed\parencite{rocklin2015dask}, running on a cluster of Raspberry Pis managed by SLURM \parencite{yoo2003slurm}.

\subsection{Global Usage Estimation}
\label{sec:usage_estimation}

We correct post count values for each country by using our country coverage estimates from Section \ref{sec:coverage} to obtain regional coverage estimates, and applying regional coverage estimates to each country in that region (for example, correcting post counts for New Zealand and other Pacific countries by assuming the same coverage as for Australia).

\subsection{Videos}

\subsubsection{Topics}

We embedded each video clip in a latent space using the \textit{xclip-base-patch32} model \footnote{https://huggingface.co/microsoft/xclip-base-patch32} \parencite{ni2022expanding}.

We then use the BERTopic \footnote{https://maartengr.github.io/BERTopic/index.html} \parencite{grootendorst2022bertopic} library to cluster the video embeddings, then use the \textit{InternVL2\_5-2B} \footnote{https://huggingface.co/OpenGVLab/InternVL2\_5-2B} \parencite{chen2024expanding} vision-language model to obtain captions for representative documents from each cluster, and finally \textit{Phi-4-mini-instruct} \footnote{https://huggingface.co/microsoft/Phi-4-mini-instruct} \parencite{abdin2024phi} to summarize clusters. We then use \textit{datamapplot} \footnote{https://datamapplot.readthedocs.io/en/latest/} to plot the resulting clusters and descriptions.

\subsubsection{Classification of posts featuring children}
\label{sec:child_classification}

We needed to build a classifier to classify each video in our 24 hour dataset as containing a child or not. The xclip video embeddings were used as the representations of each video. To train and test a classifier, we used a dataset of TikTok accounts \citep{schirmer2024more}. The dataset contains 5,248 unique posts depicting children, including babies and young teenagers below the age of 13, and an additional 6,972 posts classified as not containing children. Videos were collected through a snowball sampling approach based on child-related keywords and are not restricted to a certain language or region. We evaluated methods via F1 on a 5-fold cross-validation train/test split. We tested several classifier methods, and found a stacking ensemble method using a logistic regression, histogram-based gradient boosting classifier \footnote{https://scikit-learn.org/stable/modules/generated/sklearn.ensemble.HistGradientBoostingClassifier.html}, and an extra trees classifier \parencite{geurts2006extremely} produced the best mean F1 of 0.86.

Given the difficulty of detecting children of varying age groups in contexts, the obtained F1 score of 0.86 seems particularly dependable. We looked at false positives and true negative predictions from the classifier, and found no general trend based on demographics. As stated by \citet{schirmer2024more}, the range of contexts in which posts feature children is vast: Videos contain vlogs of new mothers showing their newborn, parents playing with their toddlers, and families performing popular TikTok dances. Further, children might be shown only for a couple of seconds or throughout the whole video. Therefore, textual descriptions of posts can contain a lot of noise or might even miss child-related cues, making the performance of our classifier even more promising.

When calculating the confidence intervals of our prediction, we wanted to account for the precision and recall of our classifier (0.88 and 0.81 respectively), the size of the test-set used to determine the accuracy of the classifier (6110), and how representative the test-set is to the real data. We quantify the representativeness of the test-set using a `data quality index' $\rho$, a number between 0 and 1 \parencite{meng2018statistical}. We set this value to 0.5, given that the videos in the test-set are from a limited range of accounts. We then calculate $N_{eff}$, the effective sample size, via the design effect, $DEFF=\frac{1}{\rho^2}, N_{eff}=\frac{N}{DEFF}$ \parencite{lohr2021sampling}. We then bootstrap 2000 samples, first sampling our precision and recall given sampling variance $prec_{sample}\sim\beta(precN_{eff}, (1-prec)N_{eff}), rec_{sample}\sim\beta(recN_{eff}, (1-rec)N_{eff})$ \parencite{walley1991statistical}. We then account for systematic bias in the test-set by adding bias via a normal distribution $prec_{bias}\sim N(prec_{sample}, 0.2(1-\rho), rec_{bias}\sim N(rec_{bias}, 0.2(1-\rho))$ \parencite{sugiyama2007covariate}. We use max $\pm20\%$ here as that is the typical performance drop under distribution shift determined by \citet{hendrycks2019benchmarking}. Finally, we simulate the classification process in terms of how many examples are mis-classified.

\section{Data Availability}

We have publicly released aggregated statistics from our collected dataset in order to preserve the privacy of those in our dataset. We have released the data at \url{https://zenodo.org/records/15330828}. We also release an explorable data map of extracted topics at \url{https://bendavidsteel.github.io/fullscreen/visual/OneHourToponymy}.

\section{Code Availability}

The code for this work is available at \url{https://github.com/bendavidsteel/tiktok-slice}


\begin{appendices}

\end{appendices}


\printbibliography

\end{document}